\documentstyle[aps]{revtex}
\tightenlines
\begin{document}
\input{epsf}
\title{Soliton Interaction  with an External Traveling Wave}
\author{Gil Cohen}
\address{Racah Institute  of  Physics, Hebrew University of Jerusalem,
Jerusalem 91904, Israel}
\maketitle
\begin{abstract}
The dynamics of soliton pulses in the Nonlinear Schr\"{o}dinger
Equation (NLSE) driven by an external Traveling wave is studied
analytically and numerically. The Hamiltonian structure of the
system is used to show that, in the adiabatic approximation for a
single soliton, the problem is integrable despite the large number
of degrees of freedom. Fixed points of the system are found, and
their linear stability is investigated. The fixed points
correspond to a Doppler shifted resonance between the external
wave and the soliton. The structure and topological changes of the
phase space of the soliton parameters as functions of the strength
of coupling are investigated. A physical derivation of the driven
NLSE is given in the context of optical pulse propagation in
asymmetric, twin-core optical fibers. The results can be applied
to soliton stabilization and amplification.
\end{abstract}
 \pagebreak

\section{INTRODUCTION} \label{I}

Investigations of the externally driven Nonlinear Schr\"{o}dinger
(NLS) equation date back to the seminal work of Kaup and Newell
\cite{KaupNewell} on the AC-driven damped NLS equation. That work
was also one of the pioneering papers in which a perturbation
method for solitons based on the Inverse Scattering Transform
(IST) technique was developed. Externally driven NLS equation
arises in many applications, mainly in the context of solid state
physics, such as long Josephson junctions \cite{josephson} and
charge density waves \cite{KaupNewell2}. The same equation
describes plasmas driven by RF fields \cite{plasma}. Much
attention has been paid to the study of chaotic phenomena in the
phase space of the soliton parameters of the driven system
\cite{Nozaki1,Nozaki2,Barashenkov}, and  to the formation and
stability of soliton states unique to the driven system
\cite{Barashenkov,Bishop}. Problems of {\it generation} of
solitons via a coupling to external perturbations have been also
investigated, for a homogeneous AC-drive \cite{lazar} and
traveling waves \cite{Baruch}.

In the model system that we shall consider (see Eq.
(\ref{SolField}) below) the perturbation term does not introduce
dissipation (the system remains conservative), and the uniform
time-dependent driving is generalized to a, spatially dependent,
driving field. The physical background of our model system is the
nonlinear pulse propagation in optical fibers. In this case the
driving force can assume, as will be shown in Sec. \ref{physical},
the form of a Traveling wave coupled linearly to the NLS equation.
The model system can be realized in the context of propagation of
envelope optical pulses in twin-core optical fibers. The aim of
this work is to study the effects of the driving field on the
dynamics of the NLS-solitons. The emphasis is on weak couplings,
and the main motivation is stabilizing and controlling solitons by
an external traveling wave.

We will limit ourselves to one-soliton pulses (that persist in an
unperturbed NLS equation) and investigate the evolution of the
parameters of the soliton when it is driven by a Traveling wave.
This form of coupling preserves the Hamiltonian structure of the
equation, thus making possible to employ  Hamiltonian perturbation
methods. We will show that, in the adiabatic approximation for a
single soliton, the problem is integrable despite the large number
of degrees of freedom. We will show that there exists a resonance,
or phase locking, between the soliton and the driving field. We
will see how this resonance is linked to the particle-like
properties of solitons and how the resonance conditions are
influenced by the strength of interaction.

Transmission of solitons over long distances is essential for the
use of solitons as digital bits in optical transmission lines
\cite{soli2}. This has been achieved by use of Iridium doped fiber
amplifiers \cite{HausRev}. Here we will suggest an alternative
scheme that employs a twin-core optical fiber. For identical
fibers with pulses centered around the same central frequency, the
twin-core system is also termed the Nonlinear Directional Coupler
\cite{nldc1}, and this system has been extensively investigated
(see \cite{myself} and references therein). Asymmetric twin-core
optical fibers were also investigated. Numerical investigation of
the asymmetric coupler as a means of performing logical gate
operations with solitons was carried out in \cite{asymmetric3}. In
\cite{asymmetric1}, ``static" soliton states which can exist in
these fibers with a limited asymmetry, were investigated in the
model of two coupled NLS equations, where new types of solitons
unique to the twin-core fiber were found.

In our model, the two coupled NLS equations, which describe the
evolution of optical pulses in asymmetric twin-core fibers, are
reduced to a single NLS equation driven by an external traveling
wave. We will see how the driving by an external wave can be used
to stabilize soliton propagation in optical fibers, suggesting
that this scheme can be used in transmission of solitons over long
distances. We will also show how solitons can be amplified by the
external Traveling wave with slowly varying parameters.

The paper is organized as follows. In Sec. \ref{Driven} we outline
the Hamiltonian perturbation method for the NLS equation driven by
an external traveling wave. In Sec. \ref{analysis} we show that
the reduced problem is integrable and investigate a novel type of
resonance (phase locking) between the soliton and the external
wave. The linear stability analysis of this resonance is
performed, and the structure and topological changes of the phase
plane are investigated, analytically and numerically. A physical
derivation of the driven NLS equation and possible applications
are presented in Sec. \ref{physical}. Sec. \ref{summary}
summarizes our results.

\section{HAMILTONIAN PERTURBATION METHOD FOR THE DRIVEN NLS EQUATION}
 \label{Driven}

Consider the NLS equation coupled to an external traveling wave
field:

\begin{equation}
i{{\partial \psi} \over {\partial t}}+{{\partial ^2\psi} \over
{\partial x^2}}+2 \left| {\psi} \right|^2\psi=\varepsilon \exp
\left[ {i \left( {kx-\omega t} \right)} \right]\;  ,
\label{SolField}
\end{equation}
where we use dimensionless variables (see Sec. \ref{physical} for the
corresponding physical
units in the case of optical fibers). In Eq. (\ref{SolField})
$\varepsilon$ is the (normalized)
strength of
the coupling, and it is assumed to be small: $\varepsilon\ll 1$. Also,
$\omega$ and $k$ are the
(normalized) frequency and wave number of the driving field,
respectively. For specific physical models $\omega$ and $k$ are related
by an appropriate dispersion relation.

The unperturbed ($\varepsilon = 0$) version of Eq.
(\ref{SolField}) is completely integrable \cite{zakharov1}, and
its most interesting solutions are solitons. The full one-soliton
solution is given by:
\begin{equation}
\psi_{\text{sol}} \left( {t,x} \right)={{{\rho \over 2}\exp \left[
{i\left( {{p \over 2}x+{{\left( {\rho ^2-p^2} \right)} \over
4}t-\hat{\varphi} -{\pi  \over 2}} \right)} \right]} \over
{\cosh\left[ {{\rho \over 2}\left( {x-pt-{{2\hat{q}} \over \rho }}
\right)} \right]}} \;, \label{unpert}
 \end{equation}
where $\rho$, $p$, $\hat{q}$ and $\hat{\varphi}$ are free
parameters. The quantity $\rho$ defines the amplitude {\it and\/}
width ($1 / \rho$) of the soliton, $p$ is the soliton's velocity,
$2\hat{q} / \rho$ is the location of its ``center of mass'', and
$\hat{\varphi}$ is its initial phase.

Eq.\ (\ref{SolField}) has Hamiltonian form, as it can be obtained
from the variational derivative of a Hamiltonian:
\begin{equation}
{{\partial \psi} \over {\partial t}}=-i{{\delta H} \over {\delta
\bar \psi }}, \label{vari}
\end{equation}
where the Hamiltonian is given by
\begin{equation}
 H\left[{\psi}\right]=\int\limits_{-\infty }^\infty  {\left( {\left|
{{{\partial \psi} \over {\partial x}}} \right|^2-\left| {\psi}
\right|^4+ 2\varepsilon \Re\left( {\bar \psi \exp \left[ {i\left(
{kx-\omega t} \right)} \right]} \right)}
 \right)}dx . \label{hamilt}
\end{equation}
Here $\bar \psi $ is the complex conjugate of $\psi$, and $\Re$
and $\Im$ are the real and imaginary parts of a complex number.

For  the unperturbed system, $\varepsilon=0$, the four parameters
of the single-soliton solution (\ref{unpert}) form a Hamiltonian
dynamical system \cite{faddeevBOOK}. For the variables
$q=\hat{q}-{{\rho p} \over 2}t$, and $\varphi=\hat{\varphi}
-{{\left( {\rho ^2-p^2} \right)} \over 4}t$, the soliton amplitude
$\rho$ becomes the canonical momentum conjugate to the coordinate
$\varphi$ while the soliton velocity $p$ is the canonical momentum
conjugate to $q$. The (reduced) one-soliton Hamiltonian of the
unperturbed system is given \cite{faddeevBOOK} by:

\begin{equation}
 H={1 \over 4}\left( {\rho p^2-{1 \over 3}\rho ^3} \right)\;. \label{H0}
\end{equation}

Going back to the perturbed system, $\varepsilon \neq 0$,
we will employ the adiabatic approximation and
investigate slow variations of the soliton parameters
caused by the external field. In doing so we neglect
radiation effects and possible formation of other solitons. This approach
is physically motivated in the context of interaction of soliton
pulses in optical fibers, where a typical initial condition is
a single soliton solution of the unperturbed NLS equation.  It is
known that solitons are
robust objects, especially in the case when the perturbations to
the integrable system are Hamiltonian \cite{Menyuk}. Therefore,
for weak couplings, it is reasonable
to expect that the main effect will be that of the solitons
persisting, but slowly changing their parameters.

It is advantageous to preserve the important Hamiltonian
properties of the perturbed problem. Therefore, we will treat the
perturbed system as a Hamiltonian system in the phase space of the
one-soliton parameters driven by the external field. The driving
term is obtained directly from the Hamiltonian (\ref{hamilt}).
Indeed, by inserting the unperturbed solution $\psi_{\text{sol}}$
with {\it time-dependent} parameters from Eq. (\ref{unpert}) into
the Hamiltonian (\ref{hamilt}), we obtain:

\begin{equation}
H\left[{\psi_{\text{sol}}}\right]=H \left({\rho, p, \varphi, q, t
}\right)= {1 \over 4} \left( {\rho p^2-{1 \over 3}\rho ^3}
\right)+\varepsilon {\pi {\sin \left[ {{{\left( {2k-p} \right)q}
\over \rho} - \omega t+\varphi } \right]} \over { \cosh\left[
{{{\left( {2k-p} \right)\pi}
 \over {2\rho}}}  \right]}}\;. \label{Hanzats}
\end{equation}

The Hamiltonian\ (\ref{Hanzats}) describes a dynamical system with
two and a half degrees of freedom, as it is nontrivially coupled
to an explicitly time-dependent driving force. As the dependence
of the Hamiltonian upon $\varphi$ and $t$ enter only through a
linear combination of these variables, a simple canonical
transformation (see below) will eliminate the explicit time
dependence, thus yielding an integral of motion: the new
Hamiltonian. The new Hamiltonian with two degrees of freedom may
still seem non-integrable, and one is tempted to look for chaos in
this system. We will show, however, that because of the existence
of an additional integral of motion, the one-soliton problem is
actually integrable and can be fully investigated analytically.
Formally, integrability occurs for {\it any} $\varepsilon$. We
expect, however, that the adiabatic approximation (that neglects
radiation and possible creation of other solitons) will be valid
only for small enough $\varepsilon$. Therefore, we will treat the
driving term perturbatively and correspondingly require that
$\varepsilon \ll 1$. On the other hand, one should assume that
$\rho > 1$, to make the contribution of the nonlinear term in the
NLS equation significant. This means that $\rho \gg \varepsilon$
which will be assumed to hold in the following.

\section{PERTURBED
ONE-SOLITON PHASE SPACE: INTEGRABILITY AND RESONANCE}
\label{analysis}

It can be easily checked that the perturbed one-soliton
Hamiltonian (\ref{Hanzats})
has an additional integral of motion,
\begin{equation}
\rho (p-2k) = \text{const}\,. \label{hint1}
\end{equation}
This integral follows from the complete, irreduced system, Eq.
(\ref{SolField}), possessing an {\it exact} conservation law:
\begin{equation} \label{int1}
 I\left[ \phi \right]=\Im \int\limits_{-\infty }^\infty
{\left( {\phi{{\partial \bar \phi} \over
{\partial x}}} \right)dx}=\text{const} \,,
\end{equation}
where $\phi= \psi \exp \left[ {-i\left( {kx-\omega t} \right)}
\right] $. Indeed, inserting the one-soliton Ansatz (\ref{unpert})
for $\psi$ in Eq. (\ref{int1}), one immediately obtains Eq.
(\ref{hint1}).

Note that the single-soliton solution form, Eq. (\ref{hint1}) in
our case, of exact integrals of the type (\ref{int1}), is only an
approximation, valid as long as generation of radiation and
formation of other solitons are ignored. Even so, it is  a good
approximation \cite{Menyuk,myself} when studying single soliton
evolution under Hamiltonian perturbations.

We now choose the integral of motion (\ref{hint1}) as a new
momentum and make the corresponding canonical transformation in
the one-soliton parameters' phase space. Simultaneously, we
exploit the abovementioned symmetry in the time dependence of the
one-soliton Hamiltonian and introduce the new phase $\Phi$ (see
below). The generating function of this canonical transformation
is:

\begin{equation}\label{gen}
  S\left({\rho,p,\Phi,Q} \right)=
-\left({p-2k}\right)\rho Q-\rho\left({\Phi+\omega t-{\pi \over
2}}\right).
\end{equation}
Therefore, the new canonical coordinates are given by

\begin{eqnarray}
&R =\rho , \qquad \qquad \qquad \qquad \qquad \qquad  &P =\rho
\left( {p-2k} \right), \nonumber
\\ &\Phi  =\varphi -{{\left( {p-2k} \right) } \over
{\rho}}q-\omega t+{\pi \over 2}, \qquad  \qquad &Q = {q/\rho}.
\label{coor1}
\end{eqnarray}
In the new coordinates the Hamiltonian becomes
\begin{equation}
 H\left({R,P,\Phi,Q} \right)=
{1 \over 4}\left[ {{1 \over R} \left( {P+2Rk} \right)^2
-{1 \over 3} R^3} \right]-\omega R +\varepsilon {\pi {\cos\Phi}
\over { \cosh \left({ {\pi P \over
{2 R^2}}} \right)}}\,, \label{hnew}
\end{equation}
and the new momentum $P$ is a constant of motion of the system.
Therefore, the Hamiltonian\ ({\ref{hnew}) represents a system with
one effective degree of freedom and is therefore integrable. This
fact excludes any possibility of chaotic motion \cite{chirikov} in
the reduced, one-soliton system.

Let us investigate possible resonances between the external wave and
the soliton. In the language of the Hamiltonian (\ref{hnew}),
an exact resonance
is related to a stable (elliptic) fixed
point. Looking for fixed points,
we should
equate $ {{\partial H}/{\partial R}}$ and $ {{\partial H}
/{\partial \Phi}}$ both to zero. This yields two conditions. The first is

\begin{equation}
\Phi_0 = \pi n \; \; \; \; \; n=0,1\,, \label{resPhi}
\end{equation}
an exact condition, valid in all orders of
$\varepsilon$. Writing down the second condition, we will first
limit ourselves to the
zero order approximation with respect to $\varepsilon$:
\begin{equation}
R_0 = \pm \left[ 2 (k^2-\omega) \pm \left( 4 (
k^2-\omega)^2 - P^2 \right)^{1/2} \right]^{1/2}\,.
\label{resR}
\end{equation}
The non-trivial, multi-valued resonance condition for the momentum
$R$ arises from the non-standard dependence of the Hamiltonian
(\ref{hnew}) on the momentum $R$.

From Eq.\ ({\ref{resR}) we obtain two conditions for the existence
of a resonance between the driving field and soliton. First, $k^2>
\omega$, which is a condition on the dispersion relation of the
external wave alone. Second, $4\left( {k^2-\omega} \right)^2 >
P^2$, which is a condition on the parameters of the soliton {\it
and} the external wave.

The linear stability of the fixed points (\ref{resPhi}) and
(\ref{resR}) is determined by the sign of the product $F G$, where
$ G = \partial^2 H_0 /
\partial R^2 (R=R_0)$  ($H_0$ is obtained from the Hamiltonian
(\ref{hnew}) by setting $\varepsilon$ to $0$), and $F=\varepsilon
\pi \cosh^{-1} \left( \pi P/ 2 R_0^2 \right)$

The phase plane of the system\ ({\ref{hnew}) for a relatively weak
coupling ($\varepsilon = 0.3$) is shown in Fig. \ref{smallE}. The
four-valued, in $R$, fixed points are clearly seen. The coordinates of the
fixed points, found numerically,
agree very well with the values given by Eqs.
(\ref{resPhi}) and (\ref{resR}).

Fig. \ref{largeE} shows the phase plane of the system
({\ref{hnew}) for a stronger coupling, $\varepsilon = 0.6$. One
can see that a topology of the phase plane has changed. Four out
of the eight fixed points have disappeared. In order to explain
this bifurcation, we should modify the resonance condition, Eq.
(\ref{resR}) and  take into account higher order corrections in
$\varepsilon$.

In the first order in $\varepsilon$ the resonance
condition for $R$ becomes:
\begin{equation}
R^4-4\left( {k^2-\omega} \right)R^2+P^2 \mp \varepsilon  {{ 4
\pi^2 P \sinh \left({ {\pi P \over {2{R_0}^2}}} \right)} \over
{{R_0} \cosh^2 \left({ {\pi P \over {2{R_0}^2}}} \right)}}=0\;,
\label{resRo1}
\end{equation}
where the values for $R_0$  should be taken form the zeroth
approximation (\ref{resR}). From Eq. (\ref{resRo1}) we can see
that indeed not all the fixed points which were present when
$\varepsilon$ was small (see Eqs. (\ref{resPhi}) and (\ref{resR}))
still persist when the coupling is increased. For a given $P$,
when

\[ \varepsilon > {{PR_0 \cosh^2 \left({ {\pi \over 2}{P \over
{{R_0}^2}}} \right)} \over {4 \pi^2 \sinh \left({ {\pi \over 2}{P
\over {{R_0}^2}}} \right)}}\;,
\]
 two of the resonant values of $R$ do
not exist anymore, thus leading to a topological change in the
phase plane, as observed in Fig. \ref{largeE}.

Now let us discuss the resonance condition in
physical terms. Transforming back to the ``lab'' coordinates, using Eq.
(\ref{coor1}), we can write the resonance condition $ \dot \Phi
= 0 $ as

\begin{equation}
  \dot{\varphi}- \frac{d}{dt} \left( {(p-2k) q \over
  R}\right)-\omega =0\; .\label{test1}
\end{equation}
Using the definition of $P$ in Eq. (\ref{coor1}), Eq.
(\ref{test1}) can be rewritten as

\begin{equation}\label{test2}
 \dot{\varphi}-  \frac{d}{dt} \left( {P q} \over
  {R^2}\right)-\omega =0\;.
\end{equation}
But $P$ is a constant of motion of our reduced system, so $
\dot{P}=0$. Also, $\dot{R}=0$ at resonance. Therefore, Eq.
(\ref{test2}) reduces to

\begin{equation}\label{test3}
\dot{\varphi}-  \left( {{ {p-2k} \over
  R}}\right) \dot{q}-\omega =0\,.
\end{equation}
Now, $ \dot {q}=R v_{s}/2$, where $v_{s}$ in the center-of-mass
velocity of the envelope of the soliton. Therefore, defining
$\kappa \equiv p/2$ as the soliton internal wave wave number (see
Eq. (\ref{unpert})), and $\omega_c \equiv \dot{\varphi}$ (in the
leading order in $\varepsilon$), the resonance condition becomes

\begin{equation}
\omega_c - \kappa v_{s} = \omega - kv_{s}\;. \label{doppler}
\end{equation}

From Eq.\ (\ref{doppler}) we can see that the resonance described
by the reduced Hamiltonian (\ref{hnew}) is a Doppler shifted
resonance between {\it two waves}: the external (pumping) wave
with the wave number $k$ and frequency $\omega$, and the carrier
wave of the soliton with the wave number $\kappa$ and frequency
$\omega_c$. It is interesting that there are two Doppler shifts in
the resonance condition. The first Doppler shift enters the right
hand side of Eq. (\ref{doppler}), and it is by the center-of-mass
velocity of the soliton. The second Doppler shift, entering the
left hand side, is less intuitive, and  is also by the
center-of-mass velocity of the soliton, but this time with the
 soliton carrier wave's wave number. This problem provides us
with another non-trivial example of particle-like properties of
solitons.

Let us look at the limiting case of $k \rightarrow 0$, in which
the system should reduce to well known resonance with a
homogeneous AC-drive \cite{KaupNewell}. For $\omega < 0$ and $p=0$
(which are the parameter values studied in \cite{KaupNewell}), the
constant of motion $P=0$ and the resonant values of the soliton's
amplitude are reduced, see Eq. (\ref{resR}), to $\rho=R=\pm
\sqrt{2\omega}$, as obtained in \cite{KaupNewell}. For $\omega >
0$, we find from Eq. (\ref{resR}) that since $k^2 > \omega$ must
hold there will be no resonance for $k=0$. So that in this case
the resonance is unique to a coupling to an external travelling
wave.

Another interesting limiting case is when the values of the wave
vectors of the driving field and the internal soliton wave are
close to each other. That is $k \simeq \kappa = p/2$. This
translates (see the transformation (\ref{coor1})) to $P
\rightarrow 0$. From Eq. (\ref{resRo1}) we can see that in this
limiting case the reduced system will be {\it exactly\/} at the
bifurcation point, for all values of $\varepsilon$. Note that for
$\kappa = 0$, also $v_s = 0$ so that the resonance (\ref{doppler})
is reduced to a simple resonance.

\section{physical model and applications} \label{physical}

In this Section we will present a physical derivation of Eq.
(\ref{SolField}). We will start with the equations for the
envelopes of pulses in the cores of two adjoining, closely spaced,
{\it non-identical\/}, single-mode fibers (twin-core optical
fibers) \cite{evanescent,asymmetric2}:
\begin{mathletters}
\label{NLDC}
\begin{equation}
 i{{\partial \psi _1} \over {\partial x}}+{{\partial ^2\psi _1}
\over {\partial t^2}}+2 \left| {\psi _1} \right|^2\psi _1+\delta
\alpha_{\text{12}} \psi _2\exp\left[ {-i \left(
{\hat{k}x-\hat{\omega} t} \right)} \right]=0, \label{NLDCa}
\end{equation}
\begin{equation}
i\left({{{\partial \psi _2} \over {\partial x}}-
{\beta_1}{{\partial \psi _2} \over {\partial t}}}
\right)+{\beta_2} {{\partial ^2\psi _2} \over {\partial t^2}}+2
\left| {\psi _2} \right|^2\psi _2+{\alpha_{\text{21}} \over
\delta} \psi _1 \exp\left[ {i \left( {\hat{k}x-\hat{\omega} t}
\right)} \right]=0. \label{NLDCb}
\end{equation}
\end{mathletters}
The coordinates $x$ and $t$ in Eqs. (\ref{NLDC})  are written in
the
``soliton units" \cite{solunits} corresponding to Eq. (\ref{NLDCa}). The
couplings $\alpha_{\text{12}}$ and $\alpha_{\text{21}}$ result from
the overlapping of the evanescent fields of the transverse fiber
modes with the fields in the adjoining fiber cores. Since the
fibers are not identical, the coupling is not symmetric, i.e.
$\alpha_{\text{12}} \neq \alpha_{\text{21}}$. It is assumed that
the transverse fiber mode is not affected by the proximity of the
adjoining fiber, and by the identical transverse
mode in it. We also assume that the interaction term arising from
the cross phase modulation (term proportional to $\left| {\psi
_i} \right|^2\psi _{3-i} , i=1,2$) can be neglected. $\delta
= \left({{\gamma_1 / \gamma_2}}\right)^{1/2}$  is
the ratio of the nonlinearity strengths in the two fibers, where
\cite{soli1}

\begin{equation} \label{gamma}
\gamma_i={{n_2\omega_i} \over {cA^{\text{eff}}_{\text{i}}}}\,\quad
\text{i}=1,2 \,,\end{equation}
$A^{\text{eff}}_{\text{i}}$ is the
effective core area (which scales like $\rho^2_{\text{i}}$,
$\rho_{\text{i}}$ being the fiber core radius), $n_2$ is the Kerr
coefficient, $c$ is the speed of light, and $\omega_i$ is the carrier
frequency in each fiber. The amplitudes $\psi _i$
of the pulses are scaled, following \cite{solunits}, to
$\left({{\gamma _i/ \beta}}\right)^{1/2}T_0$, where $\beta$
is the dispersion coefficient of the pulses in Eq. (\ref{NLDCa}),
and $T_0$ is the pulse width.

It follows from Eq. (\ref{gamma}) that the inequality $\delta \neq
1$ may result from the fibers having different radii, in which
case $\delta$ is the ratio between the radii of the fiber cores of
the two fibers. Also, if the fibers are centered around different
central frequencies $\omega_i$, then $\delta=\left({\omega_1 /
\omega_2} \right)^{1/2}$. Let us continue our discussion of the
different coefficients in Eq. (\ref{NLDC}). The coefficient
$\beta_1$ is a measure of the difference in the group velocity
 in Eq. (\ref{NLDCb}) from that in Eq.
(\ref{NLDCa}). The coefficient $\beta_2$ is the ratio of the
dispersion coefficients of the two fibers. $\beta_2 \neq 1$ may
result from the fibers not having the same transverse wave
numbers, or from the pulses in the fibers being centered around
different central frequencies. The different carrier frequencies
and/or transverse wave numbers also lead to the fibers having
different phase velocities. This fact results in the oscillatory
term in the interaction, with $\hat{k}$ and $\hat{\omega}$ being
the mismatches in the wave-number and frequency, respectively.
This oscillatory term arises in the evaluation of the overlap
integral of the transverse modes of the two fibers. Notice that,
as the coupling in Eqs. (\ref{NLDC}) is asymmetric, it is not
possible in general to cast the system in a Hamiltonian form as it
was done in \cite{asymmetric2}. We will show below, however, that
in a certain limit the system (\ref{NLDC}) can be reduced (see Eq.
(\ref{reduced})) to an equation possessing Hamiltonian structure.

We assume that the interaction term (the r.h.s.) in Eq.
(\ref{NLDCa}) is much larger than the interaction term in Eq. (\ref{NLDCb}):
$\delta\alpha_{\text{12}} \gg \alpha_{\text{21}}/\delta$.
This condition implies that $\delta \gg 1$ which occurs when the ratio
between the radii of the two fibers is large, and when the carrier
frequencies are not the same. Under these conditions
the interaction term in Eq. (\ref{NLDCb}) can be neglected,
and Eq. (\ref{NLDCb}) is decoupled from Eq. (\ref{NLDCa}) in
the sense that it only enters as a driving term in Eq.
(\ref{NLDCa}), while there is no back action.
Now, if we further assume that the pulses
described by Eq. (\ref{NLDCb}) are in the {\it positive} (normal)
dispersion regime then there is no modulational instability
\cite{soli1} in Eq. (\ref{NLDCb}), and stable linear dispersive
waves can propagate in the fiber. We are interested in the
small amplitude limit of Eq. (\ref{NLDCb}),
when the pulses are just linear waves. In this case we can drop the term
arising from the Kerr nonlinearity. Then the set of equations (\ref{NLDC}) can
be written as:

\begin{mathletters}
\label{tt}
\begin{equation}
i{{\partial \psi _1} \over {\partial x}}+{{\partial ^2\psi _1}
\over {\partial t^2}}+2 \left| {\psi _1} \right|^2\psi
_1+\varepsilon \psi _2\exp\left[ {-i \left( {\hat{k}x-\hat{\omega}
t} \right)} \right]=0\;, \label{tta}
\end{equation}
\begin{equation}
i\left({{{\partial \psi _2} \over {\partial x}}-
{\beta_1}{{\partial \psi _2} \over {\partial t}}}\right)+{\beta_2}
{{\partial ^2\psi _2} \over {\partial t^2}}=0\;, \label{ttb}
\end{equation}
\end{mathletters}
where $\varepsilon=\delta \alpha_{12}$. Therefore the system
(\ref{NLDC}) reduces to a single NLS equation driven by an
external Traveling wave.

The equation for $\psi_1$, Eq. (\ref{tta}), can be written
(omitting the indices) as:
\begin{equation}
i{{\partial \psi} \over {\partial x}}+{{\partial ^2\psi} \over
{\partial t^2}}+2 \left| {\psi} \right|^2\psi+\varepsilon \exp
\left[ {i \left( {Kx-\Omega t} \right)} \right]=0\;  ,
\label{reduced}
\end{equation}
where the dispersion relation $\Omega \left({K} \right)$ is given
by the equation for $\psi_2$ in Eq. (\ref{ttb}) together with the
phase mismatch $\hat{k}$ and $\hat{\omega}$ from Eq. (\ref{NLDC}).

Eq. (\ref{reduced}) is equivalent to Eq. (\ref{SolField}) with $x$
and $t$ interchanged, $K=-\omega$, and $\Omega=-k$. Therefore we can
apply the results obtained in Sec. \ref{analysis} to the analysis of
the dynamics of soliton pulses in twin-core fibers, under the
conditions that lead to Eq. (\ref{reduced}).

One application of our results is soliton phase locking. The
resonance conditions [see Eqs. (\ref{resPhi}) and (\ref{resR})]
correspond to the soliton's parameters being phased locked to the
driving field. For the stable fixed points this allows fixing of
the soliton parameters (see also Fig. \ref{smallE}). Specifically,
the soliton amplitude $\rho$, see Eq. (\ref{unpert}), is constant.
The values of the soliton parameters for which the resonance
condition is satisfied are given, using Eq. (\ref{coor1}), by Eqs.
(\ref{resPhi}) and (\ref{resR}), and by the value of the constant
of motion, Eq. (\ref{hint1}), which is determined by the initial
conditions of the soliton pulse and by the parameters of the
driving field. Furthermore, the resonance is sustained, as can be
seen from Fig. \ref{largeE}, also for large values of coupling
($\varepsilon$ in Eq. (\ref{SolField})), only that the number of
resonances is decreased (see Eq. (\ref{resRo1})).

The phase locking can be used in order to stabilize solitons. In
any real soliton transmission system there exists dissipation due
to fiber losses. The dissipation will result in a decrease of the
solitons' amplitude. The dissipation can be incorporated into the
NLS equation  by the addition \cite{soli1} of a perturbation term
in the form $-i\Gamma \psi$. Although the dissipative term cannot
be directly included in our Hamiltonian perturbation approach, we
can anticipate, in analogy to the problem of a driven damped
oscillator, that by keeping the soliton's parameters in resonance
with the driving field we will be able to overcome the effect of
the losses in the system. In this case, the elliptic fixed points
of Figs. \ref{smallE} and \ref{largeE} will become attracting
points.

Another application for which the coupled system can be used is
soliton amplification. The amplitude ($\rho$ in Eq.
(\ref{unpert})) of the soliton can be increased for pulses which
parameters correspond to periodic orbits surrounding stable fixed
points (see Figs. \ref{smallE} and \ref{largeE}). The period of
oscillations for the stable orbits is long and, for small
oscillations, is given by $\varepsilon^{1/2}F G$, where $F$ and
$G$ were defined following Eq. (\ref{resR}). By a proper choice of
the initial conditions and interaction length the soliton
amplitude can be increased by performing a half period of
nonlinear oscillation around a stable fixed point.

Another mechanism, by which the amplitude of solitons can be
increased more significantly, is the ``Dynamic Autoresonance''
\cite{lazar,Baruch,auto}. The frequency of nonlinear oscillations
depends on their amplitude. Therefore, the infinite growth of the
amplitude, obtained for a linear, dissipation free oscillator
driven by a resonant external force, is not possible for a
nonlinear oscillator with constant parameters. By ``chirping''
adiabatically the driving field's frequency, one can preserve the
phase plane structure so that the oscillator will continue to
perform nonlinear oscillations around the (time dependent)
resonant value of the action variable, $R$ in Eq. (\ref{hnew}) in
our case. The soliton parameters will remain phase locked in
resonance with the (slowly varying) driving field. Since $R$ is
the soliton amplitude, see Eq. (\ref{coor1}), this mechanism
provides a means for substantially increasing the soliton
amplitude in a resonant manner.

\section{Summary} \label{summary}

We have investigated the evolution of single-soliton pulses of the
Nonlinear Schr\"{o}dinger (NLS) equation driven by an external
traveling wave field. This system, even though not integrable, is
still Hamiltonian. Using the Hamiltonian structure and adiabatic
approximation for a single soliton, we reduced the perturbed NLS
equation to a two-and-a-half dimensional Hamiltonian system in the
phase space of the parameters of the single soliton solution. One
integral of motion in this system results from the fact that the
time-dependence drops out in a rotating reference frame. An
additional integral of motion is a consequence of an exact
integral of motion in the complete, unreduced partial differential
equation. Therefore, the reduced system becomes effectively
one-dimensional and therefore integrable. Physically, the reduced
system represents a nonlinear oscillator (with an unusual form of
the Hamiltonian) coupled to an external harmonic force. The phase
plane of this system was investigated analytically and
numerically, and a good agreement between these two was found. As
the coupling strength increases, there occurs a bifurcation in the
phase plane of the reduced system. This bifurcation has been
explained analytically.

We gave a physical motivation to this model by showing that
dynamics of soliton pulses in twin-core, {\it non-identical},
single-mode optical fibers can be reduced to a system in which the
dynamics in one core are governed by a NLS equation driven by a
linear Traveling wave propagating in the adjoining core. In this
regime, one can neglect the back action of the nonlinear wave on
the linear wave.

Finally, we discussed possible applications of our results to
stabilization and amplification of soliton pulses in the
asymmetric, twin-core optical fiber.

\acknowledgments

We are extremely grateful to B. Meerson for a continuous interest
and advice, and for a critical reading of the manuscript.

\begin{figure}
\caption{The phase plane of the Hamiltonian system
(\protect\ref{hnew}) for a weak coupling.
The parameters are $\varepsilon=0.3, P=
3.0$, $k=2.0$ and $\omega=2.0$ . The resonances corresponding to
the four-valued solution of Eq. (\protect\ref{resR}) are clearly
seen. Note that the stability of the resonant points is
interchanged for positive and negative values of $R$. }
\label{smallE}
\end{figure}
\begin{figure}
\caption{The phase plane of the Hamiltonian system, Eq.
(\protect\ref{hnew}), for a stronger coupling. $\varepsilon=0.6$,
The parameters are the same as in Fig. \protect\ref{smallE},
except that $\varepsilon=0.6$.  A bifurcation has occurred. In
accordance with Eq. (\protect\ref{resRo1}), for each of the
resonant values of $R$ there are no longer two resonant phases,
one for a stable (elliptic) point and one for an unstable
(hyperbolic) point. Only one fixed point is left for each value of
$R$. The phase plane topology has changed accordingly.}
 \label{largeE}
\end{figure}

\end{document}